\documentclass{article}
\usepackage[utf8]{inputenc}
\usepackage{amssymb}
\usepackage{amsfonts}
\usepackage{amsmath}
\usepackage{mathtools}
\usepackage{commath}
\usepackage{array}
\usepackage{esvect}
\usepackage{cite}
\usepackage[dvipsnames]{xcolor}
\usepackage{physics}
\usepackage{mathrsfs}
\usepackage{authblk}
 \usepackage{multirow}
\usepackage{geometry}
\usepackage{abstract}
\usepackage[english]{babel}
\usepackage[T1]{fontenc}
\usepackage[colorlinks,urlcolor=blue,citecolor=blue]{hyperref}
\begin{document}

%

\begin{center}
{\LARGE \bf
Integrable extensions of two-center Coulomb systems\\
}
\vspace{6mm}
{\Large Francisco Correa$^a$, Octavio Quintana$^{b}$
}
\\[6mm]
\noindent ${}^a${\em 
Instituto de Ciencias F\'isicas y Matem\'aticas\\
Universidad Austral de Chile, Casilla 567, Valdivia, Chile}\\ 
 {\tt francisco.correa@uach.cl}\\[3mm]
\noindent ${}^b${\em
Facultad de Física, Pontificia Universidad Cat\'olica de Chile,   \\
Vicu\~na Mackenna 4860, Santiago, Chile} \\
 {\tt oaquintana@uc.cl}
\vspace{8mm}
\end{center}

\begin{abstract} In this paper, we investigate new integrable extensions of two-center Coulomb systems. We study the most general $n$-dimensional deformation of the two-center problem by adding arbitrary functions supporting second order commuting conserved quantities. The system is superintegrable for $n>4$ and, for certain choices of the arbitrary functions, reduces to known models previously discovered. Then, based on this extended system, we introduce an additional integrable generalisation involving Calogero interactions for $n=3$. In all examples, including the two-center problem, we explicitly present the complete list of Liouville integrals in terms of second-order integrals of motion.
\end{abstract}

\section{Introduction}
The study of the hydrogen atom system, often referred to as the Coulomb problem, is a focal point in quantum mechanics because of its widespread presence and applications in nature, as well as its valuable historical background.  The Coulomb problem also exhibits quantum integrability and maximal superintegrability, reasons behind its ``accidental'' degeneracy \cite{cqs, Pauli}. The latter arises from the presence of a non-trivial integral of motion, the Laplace-Runge-Lenz vector operator, see \cite{cqs, cordani, cariglia} and references therein. Extending this investigation to $n$ dimensions, classical and quantum maximally superintegrable properties have been established \cite{Giorgi, Alliluev}.

One of the first explorations of multiparticle systems in quantum mechanics was the analysis of the ionised hydrogen molecule, which is the simplest form of a diatomic molecule and a fundamental case of the two-center Coulomb problem \cite{dmca, Teller}. This system is the quantum version of the Euler's three-body problem studied in 1760 \cite{cariglia, mat}. In the first half of the 20th century, the separability of the Schrödinger equation in elliptic coordinates for the two-center problem was already known. Erikson and Hill \cite{Erikson} presented the quantum mechanical version of an extra integral of motion inspired from the Laplace-Runge-Lenz vector for the case $n=3$, showing its intrinsic connection to the separability of the system in prolate spheroidal coordinates. Coulson and Joseph \cite{Coulson} extended this work to arbitrary dimensions, showing that the form of the non-trivial integral of motion remains consistent in Euclidean space. However, in the presence of more than two centers the system ceases to be integrable, since the symmetries which allow the existence of a non-trivial conserved quantity derived from the Laplace-Runge-Lenz vector operator disappear \cite{Coulson}. The integrable properties of the two-centre Coulomb problem have been studied from various approaches, including for instance exact and quasi-exact solvability or supersymmetric quantum mechanics  \cite{ kmp, thep1, gmt, susy1, ko, elem1}. 

In a series of papers, Helfrich, Hartmann and Kehl \cite{Helfrich-Hartmann, Kehl, Helfrich}, worked on generalised diatomic orbitals, introducing interactions dependent on the inverse of the product of the radii in three dimensions. Unlike other interactions, this extension preserved the rotational symmetry of the two-center problem, maintaining integrability and revealing a set of conserved quantities. The separability and the existence of more general integrable three dimensional extensions depending on arbitrary functions were studied by Miller Jr. and Turbiner  \cite{miltur}. The separability of the two-center problem in higher dimensions has been studied extensively in \cite{separability}, using hyperspheroidal coordinates for $n>3$. Recent extensions of the two-center problem, such as those explored in \cite{TA, TA2}, have considered a Calogero type of interaction, leading to partial separability in hyperspheroidal coordinates and the construction of the corresponding non-trivial integral of motion using Dunkl operators \cite{dunkl}. 

Considering the previous integrable deformations of the two-center problem, we can ask how general an interaction depending on both radii can be such that it is possible to construct a deformation of the quadratic integral of motion discovered by Erikson and Hill and thus separability of the Schrödinger equation. In the present work we find two families of interactions which allow the existence of a non-trivial conserved quantity, including the most general radii-only interaction in $n$ dimensions which allows separability in hyperspheroidal coordinates. We also explicitly show the $n-2$ quadratic conserved quantities under involution in terms of the angular momenta, filling a gap in the literature on the integrability in the higher-dimensional two-center Coulomb problem. 

The paper is organised as follows. In Section \ref{sec2} we review the main integrable algebraic properties of the central problem in $n$ dimensions and the special case of the Coulomb potential. This basis will prove useful for the comparison with the different models that will be studied next. Section \ref{sec3} is devoted to the summary of the algebraic structure of the $n$-dimensional two-center Coulomb problem, including the explicit form of the complete set of conserved quantities. In Section \ref{sec4} we introduce a generic extension of the $n$-dimensional two-center Coulomb and study the integrable features and the separability of the corresponding Schrödinger equation. As an application of this scheme, in Section \ref{sec5} we construct a further extension involving Calogero type of interactions for three dimensions (or three particles) with second order charges. In Section \ref{sec6} we discuss open problems and final remarks.

\section{Superintegrability of the $n$-dimensional central \& Coulomb problems} \label{sec2}

Let us consider a $n$-dimensional particle with coordinates $\Vec{x}=(x_1,x_2,\ldots,x_{n})$ and momentum operators ${\Vec{p}}=(p_1,p_2,\ldots,p_n)$.  Setting the natural units of $\hbar=m=1$, we have $p_{\ell}=-i\partial_\ell$ and $[x_{i},p_{j}]=i\delta_{ij}$, where $\delta_{ij}$ is the Kronecker delta. As a starting point, we choose a generic central potential $V(r)$, which depends only on the radial coordinate $r=\sqrt{x_1^2{+}x_2^2{+}\ldots{+}x_n^2}$. The spherically symmetric Hamiltonian operator,
\begin{equation}\label{coulomb}
    H=\frac{1}{2}\vec{p}\,^2+V(r) \, ,
\end{equation} 
remains invariant under $SO(n)$ rotations about the origin. Their associated conserved quantities, $[H,L_{ij}]=0$, are the components of the angular momentum tensor defined by
\begin{equation}\label{angmom}
    L_{ij}=x_ip_j-x_jp_i \ ,\  i,j=1,2,\ldots,n \, ,
\end{equation} 
that satisfy the commutation relations of the standard $SO(n)$ generators, 
\begin{equation}
\label{commuang}
     [L_{ij},L_{kl}]=i(\delta_{jl}L_{ik}+\delta_{ik}L_{jl}-\delta_{jk}L_{il}-\delta_{il}L_{jk}) \, .
\end{equation}
Following the notion of quantum integrability \cite{cqs}, we look for $n$ commuting conserved quantities. From the commutation relations (\ref{commuang}), it is clear that we can choose, for example, the set of operators (\ref{angmom}) such that $i \neq j\neq k\neq l$. However, it is possible to find a more convenient set by constructing the quadratic Casimir elements of the corresponding groups $SO(2)\subset \cdots SO(k)\subset \cdots \subset SO(n)$ \cite{TA},  
\begin{equation}\label{casimirs}
C_{k}=\sum_{i<j}^k L_{ij}^2 \ , \ \ \ 2\leq k \leq n \, , 
\end{equation}
where $C_{n}=\vec{L}^2$. By construction, these operators are in involution, i.e., they satisfy $[C_{i},C_{j}]=0$, for any $2\leq i<j\leq n$. In this way, the quantum analog of Liouville integrablity is given in terms of the following conserved charges,
\begin{equation}
\text{Liouville integrals:} \quad \left\{H,C_{2}, C_{3}, \ldots, C_{n} \right\} \, .
\end{equation}
As one of the standard examples of superintegrable systems, the $n$ dimensional central problem (\ref{coulomb}) at classical and quantum level contains, in addition of the Liouville integrals, extra conserved quantities. However, among the $n(n-1)/2$ components of the angular momentum tensor (\ref{angmom}), there are only $2n-3$ independent ones as a consequence of the identity
\begin{equation}\label{relA}    L_{ij}L_{kl}+L_{il}L_{jk}+L_{ik}L_{lj}=0 \, , \quad i \neq j\neq k\neq l \, .
\end{equation}
for $i,j,k,l$ distinct. This can be seen, for instance, choosing $i=1, \ j=2$ in the above relation, $L_{12}L_{kl}+L_{1l}L_{2k}-L_{1k}L_{2l}=0$, where $3\leq l<k\leq n$ such that every component $L_{kl}$ can be expressed only in terms of the $2n-3$ components $\{ L_{12},\ldots,L_{1n},L_{23},\ldots,L_{2n}\}$. In this way, together with the Hamiltonian, the system possesses a total of $2n-2$ independent integrals of motion, which establishes the superintegrability of the central problem. Further details on the independence of the angular operators in the classical case can be found in \cite{Central}. 
\subsection{Coulomb case}
As one of the standard problems in quantum mechanics, the hydrogen atom or Coulomb potential  
\begin{equation}\label{coum}
 V(r)=-\frac{\gamma}{r}
\end{equation}
has been studied extensively over the last century, finding many applications from group theory to black hole physics \cite{cordani, cariglia, Sakurai, bhsym}. Here we will briefly review some details of the algebraic structure of its integrability in $n$ dimensions. Besides the charges given by the angular momentum operators discussed in the previous case for generic central potentials, the not-so-accidental degeneracy of the spectrum is explained by the Laplace-Runge-Lenz vector. Using this conserved operator, Pauli first solved the spectrum for the 3D case in 1926 \cite{Pauli}, before the Schrödinger equation was known. The choice of the potential (\ref{coum}) endows the system with $n$ additional conserved quantities, which are the components of the quantum Laplace-Runge-Lenz vector \cite{Giorgi}, 
\begin{equation}\label{LRL}
    A_{i}=\frac{1}{2}\sum_{j=1}^{N} (L_{ij}p_{j}+p_jL_{ij})-\gamma\frac{ x_i}{r} \,  , \quad [A_{i}, H]=0\, ,\quad i=1,\ldots,n.
\end{equation}
The algebra between this operator and the angular momentum (\ref{angmom}) takes the following form 
\begin{equation}\label{lrlalgebra}
[A_i, A_j]=-2 i H L_{ij}, \quad [A_i,L_{jk}]=i (\delta_{ik} A_j-\delta_{ij} A_k) \, ,
\end{equation}
where we recognise its non-linear nature. The discussion of how this algebra (\ref{lrlalgebra}) can be linearised for fixed energy levels explaining the spectrum degeneracy, is studied in standard quantum mechanics textbooks, see for example \cite{Sakurai}. From the second commutation relations in (\ref{lrlalgebra}), we can check that some of the components of the $A_i$ vector will not be completely independent when the $L_{ij}$ operators are taken into account. In the three dimensional case, this fact is expressed as the constraint $\vec{A} \cdot \vec{L}=0$. In addition, from the non-linear relationship,
\begin{equation}\label{aec}
\vec{A}\,^2=H \left(2 \vec{L}\,^2+ {(n-1)^2} \right)+\gamma^2 \, ,
\end{equation}
it can be seen that in the end the $n$-dimensional Laplace-Runge-Lenz vector adds only an independent integral of motion, so that the Hamiltonian (\ref{coulomb}) with Coulomb potential (\ref{coum}) becomes maximally superintegrable.

\section{Integrability of the $n$-dimensional two-center Coulomb problem} \label{sec3}
The next case we consider is a $n-$dimensional charged particle under the influence of two charged centers \cite{dmca, Teller, Erikson, Coulson} described by the Hamiltonian,
\begin{equation}
\label{121}
    H_{\text{2c}}=\frac{1}{2}\vec{p}\,^2-\frac{\gamma_+}{r_+} -\frac{\gamma_-}{r_-}\, ,
\end{equation}
where $r_\pm$ are the distances from the charged moving particle to two fixed centers. These centers are chosen to be located at the positions $\vec{{a}}=a\hat{u}$ and $-\vec{a}$, along a line defined by the direction\footnote{The problem can also be defined in such a way that the centers of charges lie on one of the $n$ basis axes \cite{Coulson, Erikson, separability}. The  definition (\ref{122}) is useful because it ensures the permutation invariance in the position coordinates of the Hamiltonian $H_{\text{2c}}$.},
\begin{equation}\label{122}
    {\hat{u}}=\frac{1}{\sqrt{n}}(1,1,\ldots,1) \, ,
\end{equation}
such that the distances $r_\pm$ are written as,
\begin{align}\label{123}
 r_{\pm}=\abs{\vec{{x}}\pm \vec{{a}}}=\Bigg[\sum_{i=1}^n \left( x_{i}\pm\frac{a}{\sqrt{n}}\right)^2\Bigg]^{1/2}=\sqrt{r^2+a^2\pm2(\vec{{a}}\cdot\vec{{x}})} \, .
\end{align}
The Hamiltonian (\ref{121}) is no longer invariant under $SO(n)$ rotations around the origin, but invariant under $SO(n{-}1)$  rotations around the line connecting  the two charges (\ref{122}). One can then write the new conserved quantities, let us call them $L_{ij}^\perp$, in terms of the previous angular momentum operators $L_{ij}$ from Eq. (\ref{angmom}), which are no longer conserved. The new integrals, $[H_{\text{2c}},L_{ij}^\perp]=0$,  are orthogonal to the $n$-dimensional vector $\hat{u}$ with components $u_k$ reading  \cite{TA, TA2}, 
\begin{equation}\label{124}
     L_{ij}^{\perp}=L_{ij}+\sum_{k=1}^{n}(u_i u_k L_{jk}-u_j u_kL_{ik})=L_{ij}+\frac{1}{n}\sum_{k=1}^n(L_{jk}-L_{ik})\, , \quad 
     1 \leq i<j \leq n \, .
 \end{equation}
Their symmetry algebra is analogous to (\ref{commuang}), but slightly different, 
  \begin{equation}\label{333}
 \quad [L_{ij}^{\perp},L_{kl}^{\perp}]=i(\tilde{\delta}_{jl}L_{ik}^{\perp}+\tilde{\delta}_{ik}L_{jl}^{\perp}-\tilde{\delta}_{jk}L_{il}^{\perp}-\tilde{\delta}_{il}L_{jk}^{\perp}) \, , \quad \tilde{\delta}_{ij}=\delta_{ij}-\frac{1}{n} \, .
\end{equation}
Following the same previous discussion in the central problem, since  the symmetry corresponds to the $SO(n{-}1)$ group, it is possible to write down $n-2$ quantities quadratic in $L_{ij}^\perp$ in involution analogous to (\ref{casimirs}),
\begin{equation}
 \label{casimirs2}
     C_{\alpha}^{\perp}=\sum_{i<j}^{\alpha}L_{ij}^{\perp^2} +\frac{1}{\alpha}\sum_{i,j,k}^{\alpha}L_{ij}^{\perp}L_{jk}^{\perp} \ , \quad \ \alpha=3,\ldots, n \, .  
 \end{equation}
Although the existence of these $n-2$ quantities quadratic in $L_{ij}^\perp$ was mentioned in \cite{TA2}, to our knowledge they have not been constructed explicitly in the literature before. The independence of the integrals (\ref{casimirs2}) can be understood easily from the fact that they can be written as $C_{\alpha}^{\perp}=\sum_{i<j}^{\alpha}L_{ij}^2 +\frac{1}{\alpha}\sum_{i,j,k}^{\alpha}L_{ij}L_{jk}$ with $\alpha=3,\ldots, n$. To this extent, the rotational symmetry from the charges (\ref{casimirs2}) and the Hamiltonian form a set of $n-1$ integrals of motion in involution, but they are not enough to trigger integrability.   

Remarkably, this system also inherits a non-trivial conserved quantity from the Laplace-Runge-Lenz vector of the hydrogen atom/Coulomb problem, first constructed in \cite{Erikson} for the three-dimensional case. This non-trivial integral was then obtained in the $n$-dimensional case by an elegant trick \cite{Coulson}, translating the projection of the Laplace-Runge-Lenz vector onto $\hat{u}$, leading to the second order operator
\begin{equation}\label{2cintegral}
A_{\text{2c}}=\vec{L}\,^2+(\vec{{a}}\cdot\vec{{p}})^2-2(\vec{{a}}\cdot\vec{{x}})\bigg(\frac{\gamma_{-}}{r_{-}}-\frac{\gamma_{+}}{r_{+}} \bigg) \, .
\end{equation}
Besides commuting with the Hamiltonian (\ref{121}), $[A_{\text{2c}},H_{\text{2c}}]=0$, it does also for the angular operators (\ref{124}), $[A_{\text{2c}},L_{ij}^\perp]=0$, and thus, with the $n-2$ mutually commuting integrals of motion (\ref{casimirs2}). In this way, we have a set of $n$ conserved quantities in involution and the system described by the Hamiltonian (\ref{121}) is integrable,
\begin{equation}
\text{Liouville integrals:} \quad \left\{H_{\text{2c}},A_{\text{2c}}, C_{3}^{\perp}, \ldots, C_{n}^{\perp} \right\} \, .
\end{equation}
Since the integrability of the two-center problem is given by conserved quantities of second order, its Schrödinger equation is separable in hyperspheroidal coordinates. This fact has been studied in \cite{separability}, together with the solutions for the $n$-dimensional case, expressed in terms of confluent Heun functions. We will study the separability of extended two-center systems in section \ref{sec4}, which includes this case as a special limit.

In order to establish the superintegrable nature of the two-center problem, we need to find out the independence of the charges $L_{ij}^{\perp}$ and $C_{\alpha}^{\perp}$, similar to what was done in section \ref{sec2}. In this case, instead of the relation (\ref{relA}), the operators $L_{ij}^{\perp}$ satisfy the following identity,
\begin{equation}    L_{ij}^{\perp}L_{kl}^{\perp}+L_{il}^{\perp}L_{jk}^{\perp}+L_{ik}^{\perp}L_{lj}^{\perp}=\frac{i}{n}(L_{jk}^{\perp}+L_{kl}^{\perp}+L_{lj}^{\perp}) \, , \quad i \neq j\neq k\neq l \, .
\end{equation}
Once we fix two indexes $i,j$, any other element $L_{kl}^{\perp}$ can be determined in terms of other components containing either $i$ or $j$ in their entries. For example, we can choose $i=1$ and $j=2$, and the elements from the sets $\{ L_{1k}^{\perp}, k=2,\ldots,n\}$ and $\{ L_{2j}^{\perp}, j=3,\ldots,n\}$ can be used to write any other element $L_{jk}^{\perp}$ with $j,k\neq 1,2$. In principle, we can regard these elements as part of a set of independent angular momentum components, but the existence of the additional constraint,
\begin{equation}
    \sum_{j=1}^n L_{ij}^{\perp}=0 \, ,
\end{equation}
reduces the set of independent orthogonal angular momenta to the set of $2n-5$ components $\{ L_{13}^{\perp}, \ldots, L_{1n}^{\perp}\}$ and $\{L_{24}^{\perp},\ldots L_{2n}^{\perp} \}$. Together with the Hamiltonian and the non-trivial integral (\ref{2cintegral}), we have a set of $2n-3$ independent integrals of motion and the superintegrability of the two-center problem is thus obtained for $n>3$.

 \section{Integrability of two-center problem with an arbitrary extension}\label{sec4}
Several deformations of the two-center problem have been studied in the literature, such as the harmonic oscillator deformation \cite{Coulson} or the extra term introduced by Helfrich and Hartmann \cite{Helfrich-Hartmann, Kehl}. In this sense, one may wonder what kind of interaction terms could be added to the two-center problem in order to preserve a conserved quantity of the type $A_{\text{2c}}$ and integrability. Although the three dimensional case was considered in \cite{miltur}, the answer to this question in $n$ dimensions can be given by considering the following Hamiltonian with only radii-like interactions, 
 \begin{equation}\label{hamilf}
H_{F}=\frac{1}{2}\vec{p}\,^2-\frac{\gamma_-}{r_-}-\frac{\gamma_+}{r_+}-\frac{1}{r_- r_+}\left(F_\mu
+F_\nu\right) \, ,
\end{equation}
where the functions $F_{\mu}$ and $F_{\nu}$ depend only on the difference and sum of $r_\pm$,
 \begin{equation}\label{func}
F_\mu=F_\mu(r_++ r_-), \quad F_\nu=F_\nu(r_+- r_-) \, .
\end{equation}
The extended potential in (\ref{hamilf}) turns out to be the most general form for which the system inherits a non-trivial conserved quantity derived from the Laplace-Runge-Lenz vector method \cite{Erikson, Coulson}. This modification of the potential transforms the integral (\ref{2cintegral}) into a new conserved charge,
 \begin{equation}\label{inta}
   A_F=\vec{L}\,^2+(\vec{{a}}\cdot\vec{{p}})^2-2(\vec{{a}}\cdot\vec{{x}})\bigg(\frac{\gamma_-}{r_-}-\frac{\gamma_+}{r_+} \bigg) -\frac{1}{2 r_- r_+}\big[(r_+-r_-)^2 F_\mu+(r_++r_-)^2 F_\nu \big] \, .
\end{equation}
Note that any function of the radii $f(r_+,r_-)$ commutes with the components $L_{ij}^{\perp}$, which means that the orthogonal angular momentum operators are still conserved quantities and the previous discussion about the independence of quadratic ones (\ref{casimirs2}) remains the same. We have then constructed the following set of mutually commuting operators,
\begin{equation}
\text{Liouville integrals:} \quad \left\{H_{F},A_{F}, C_{3}^{\perp}, \ldots, C_{n}^{\perp} \right\} \, .
\end{equation}
In the table (\ref{table:1}) we summarise some known cases for which the generic interactions $F_\mu$ and $F_\nu$ in (\ref{hamilf}) reduce to the Helfrich and Hartmann case, harmonic oscillator and an extra combination, and the corresponding form of the integral (\ref{inta}). 
\renewcommand{\arraystretch}{2.5}
\begin{table}[h!]
\begin{center}
\begin{tabular}{c|c|c|c|c|}
\cline{2-5}
                       & $F_\mu$ & $F_\nu$ & Potential & Contributions to $A_F$ \\ \hline
\multicolumn{1}{|c|}{Helfrich-Hartmann (H{-}H)} & \textit{C=constant} & $Q-C$ &  $\displaystyle \frac{Q}{r_+ r_-}$  & $\displaystyle -Q\frac{r^2+a^2}{r_{+}r_{-}}$ \\ [1ex] \hline
\multicolumn{1}{|c|}{Harmonic oscillator} &  $-\frac{k}{2}(r_{+}+r_{-})^4$ & $\frac{k}{2}(r_{+}-r_{-})^4$ & $-\frac{k}{2}(r_{+}^2+r_{-}^2)$ & $\frac{k}{8}(r_{+}^2-r_{-}^2)^2=2k(\vec{a} \cdot \vec{x})^2$ \\ \hline
\multicolumn{1}{|c|}{H{-}H {+} Harmonic mix-up} & $-\frac{1}{4}(r_{+}+r_{-})^2$ & $-\frac{1}{4}(r_{+}-r_{-})^2$ & $\displaystyle -\frac{r^2+a^2}{r_{+} r_{-}}$ &  $\displaystyle \frac{4(\vec{a} \cdot \vec{x})^2}{r_{+}r_{-}}$ \\ [1ex] \hline
\end{tabular}
\end{center}
\caption{Reduction of the system $H_{F}$ and integral $A_F$ to the Helfrich-Hartmann and Harmonic oscillator interactions.}\label{table:1}
\end{table}
As the extensions introduced in (\ref{hamilf}) keep the orthogonal symmetries provided by $L_{ij}^{\perp}$, the superintegrability remains unaltered as in the previous section. Since the order of the commuting charges is also two, we can also expect the Schrödinger equation to be separable. We will now focus on this using hyperspheroidal coordinates.

Let us introduce the Jacobi coordinates \cite{Reed}, which describe a system with $n$ degrees of freedom by separating the center of mass coordinates $y_0$ from the relative coordinates $\vec{y}=(y_1,y_2\ldots,y_{n-1})$, while keeping the kinetic terms unchanged,\begin{equation}\label{211}
    y_0=\frac{1}{\sqrt{n}}\sum_{i=1}^n x_i , \ y_k=\frac{1}{\sqrt{k(k+1)}}\bigg(\sum_{j=1}^kx_j -kx_{k+1}\bigg) \, , \quad k=1,\ldots,n-1 \, .
\end{equation}
In these coordinates $y_0^2+{\vec{y}}^{\,2}=r^2$ and the charged centers lie in the center of mass axis $y_0$ where $r_\pm=\sqrt{y^2+(y_0\pm a)^2}$. Next, we introduce the hyperspheroidal coordinates in terms $\mu$ and $\nu$ and the $n-2$ angles $\varphi_{\ell}$ that parametrize the $S^{n-2}$ sphere \cite{separability},
\begin{equation}
\mu=\frac{r_++r_-}{2a}\, ,\quad \nu=\frac{r_--r_+}{2a}\, ,\quad   \varphi_{\ell}\, , \quad \ell=1,\ldots,n-2 \, .
\end{equation}
The relation with the Jacobi coordinates contains the unitary vector $\hat{n}(\varphi_{\ell})$ pointing radially out of the $S^{n-2}$ sphere,
\begin{equation}
    y_0=-a\mu\nu\, , \quad  \vec{y}=a\sqrt{(\mu^2-1)(1-\nu^2)} \, \hat{n}(\varphi_{\ell})\, .
\end{equation}
The Hamiltonian in terms of $(\mu,\nu,\varphi_{\ell})$ takes the following form,
\begin{multline}
H_F=-\frac{1}{2a^2(\mu^2-\nu^2)}\bigg[ \frac{1}{(\mu^2-1)^{\frac{n-3}{2}}}\partial_{\mu}(\mu^2-1)^{\frac{n-1}{2}}\partial_\mu+\frac{1}{(1-\nu^2)^{\frac{n-3}{2}}}\partial_{\nu}(1-\nu^2)^{\frac{n-1}{2}}\partial_\nu  \bigg]\\-\frac{1}{2a^2(\mu^2-1)(1-\nu^2)}\nabla^2_{S}-\frac{\gamma_+}{a(\mu+\nu)}-\frac{\gamma_-}{a(\mu-\nu)}-\frac{F_\mu+F_\nu}{a^2(\mu^2-\nu^2)}\, .
\end{multline}
Finally, we can separate the Schrödinger equation $H_F\psi=E\psi$ with the wave function ansatz depending on $\mu$, $\nu$ and the angles $\varphi_l$ in the following way,
\begin{equation}\label{ans}
\psi(\mu,\nu,\varphi_{l})=\chi_+(\mu)\chi_-(\nu)\Xi(\varphi_{\ell}) \, .
\end{equation}
The corresponding separable equations for $\mu$ and $\nu$ take a similar form,
\begin{align}\label{eqs1}
    \Bigg[&\frac{1}{(\mu^2-1)^{\frac{n-3}{2}}}\partial_{\mu}(\mu^2-1)^{\frac{n-1}{2}}\partial_\mu+2a(\gamma_+ +\gamma_-)\mu+2a^2(\mu^2-1)E-\lambda+2F_\mu-\frac{q(q+n-3)}{\mu^2-1} \Bigg]\chi_+(\mu)=0 \,  \\  \label{eqs2}
    \Bigg[&\frac{1}{(1-\nu^2)^{\frac{n-3}{2}}}\partial_{\nu}(1-\nu^2)^{\frac{n-1}{2}}\partial_\nu-2a(\gamma_+-\gamma_-)\nu-2a^2(1-\nu^2)E+\lambda+2F_\nu-\frac{q(q+n-3)}{1-\nu^2} \Bigg]\chi_-(\nu)=0 \, ,
\end{align}
depending on the separation constants $\lambda$ and $q(q+n-3)$. Note that the separation constant $\lambda$ is closely related to the eigenvalues of the integral of motion $A_F$ (\ref{inta}), as has already been observed for some special cases \cite{Helfrich, miltur}. The remaining angular equation can be written in terms of the $n-2$ dimensional spherical Laplacian $\nabla^2_{n-2}$,
\begin{equation}\label{angeq}
\nabla^2_{n-2} \Xi(\varphi_{\ell})=-q(q+n-3)\Xi(\varphi_{\ell}) \, .
\end{equation}
Writing the angular dependence as $\Xi(\varphi_{l})=\Pi_{k=1}^{n-2}\Xi_k(\varphi_{k})$, the equations completely separate into the following set of equations,
\begin{equation}\label{2213}
\bigg[ \frac{1}{\sin^{k-1}{\varphi_k}}\frac{\partial}{\partial \varphi_k} \sin^{k-1}{\varphi_k}\frac{\partial}{\partial \varphi_k}-\frac{m_{k-1}(m_{k-1}+k-2)}{\sin^2{\varphi_k}}+m_k(m_k+k-1)\bigg] \Xi_{k}(\varphi_{k}) =0,
\end{equation}
\begin{equation}
\label{2212}
\bigg[ \frac{\partial^2}{\partial \varphi_1^2}+m_1^2 \bigg]\Xi_{1}(\varphi_{1}) =0,  
\end{equation}
where $k=2,\ldots,n-2$, and the quantities $m_1, m_2,\ldots,m_{N-2}=q$ are the separation constants.  We conclude that the potential considered in (\ref{hamilf}) is the most general that achieves complete separability of the $n$-dimensional Schrödinger equation via a conserved quantity of the form $A_F$. In the next section we will use this scheme to consider a further extension via Calogero interactions. 

 \section{Integrability of two-center and Calogero extensions in $3D$} \label{sec5}
For the last example, we consider $n=3$ dimensions, which can also be thought of as three identical particles of equal mass $m=1$. Starting from the Hamiltonian $H_F$ (\ref{hamilf}), we add the pairwise inverse square interactions given by the rational Calogero model \cite{calo1, calo2, calo3} depending on a coupling parameter $g$,
 \begin{equation}\label{hamilcal}
H_g=H_F+\sum_{\ell<m}^3\frac{g(g-1)}{(x_\ell-x_m)^2} \, .
\end{equation}
The Calogero model (also known as Calogero-Moser-Sutherland) is an excellent, if not the best, example of many particle integrable systems, both at the classical and quantum level. Most of the properties of integrable models can be tested and studied in the many variants and deformations of the Calogero models. In fact, in the quantum regime, the Dunkl operators \cite{dunkl} appear as quite useful objects to understand the integrable properties of the integrals of motion, the intertwining operators and the algebraic integrability \cite{opdam, heckman, clp, ccl}. In the present case, the Hamiltonian (\ref{hamilcal}) can be obtained in terms of the Dunkl operators, which have the following form
\begin{equation}
{\cal D}_\ell=\partial_\ell +\sum_{\ell<m}^3 \frac{1}{x_\ell-x_m} s_{\ell m}, \quad {\cal P}_\ell=-i {\cal D}_\ell \, , \quad \ell=1,2,3 \, .
\end{equation}
The symbols $s_{\ell m}$ represent the permutation operators, $s_{\ell m} x_m=x_\ell s_{\ell m}$, $s_{\ell m} p_m=p_\ell s_{\ell m}$ and satisfy $s_{\ell m}^2=1$ for $\ell,m=1,2,3$. Replacing the derivatives by the Dunkl operators $p_\ell \rightarrow {\cal P}_\ell$ in the Hamiltonian $H_F$ gives a new non-local operator which explicitly depends on the permutation operators $s_{\ell m}$. Shifting the permutations to the right and then restricting the resulting operator to act on completely symmetric functions eliminates all non-local terms and reduces the Hamiltonian to the form (\ref{hamilcal}). For example, in the absence of the two-center extension, the Calogero Hamiltonian is obtained simply in terms of Dunkl operators via $\textrm{res}(\frac{1}{2}\vec{{\cal P}}\,^2)=\frac{1}{2}\vec{p}\, ^2+\sum_{\ell<m}^3\frac{g(g-1)}{(x_\ell-x_m)^2}$, where $\textrm{res}(\cdot)$ is the restriction to acting on symmetric functions. In the two-center Coulomb problem, which corresponds to $F_\mu=F_\nu=0$ in $H_g$, this idea has indeed been used by Hakobyan and Nersessian \cite{TA,TA2}. They introduce the unit vector $\hat{u}$ and the deformed momentum operators $L_{ij}^{\perp}$ with the advantage that the symmetric choice of $\hat{u}$ makes the Hamiltonian invariant under the permutation transformations given by $s_{\ell m}$, $[H_g, s_{\ell m}]=0$ for $\ell,m=1,2,3$.  

Under this symmetric scheme, it is easy to apply the same idea of replacing Dunkl operators by derivatives and subsequent reduction to the case of the integrals $C_{3}^{\perp}$ and $A_F$ in (\ref{casimirs2}) and (\ref{inta}), respectively. Note that among the different choices of $\alpha$ in (\ref{casimirs2}), only $\alpha=n=3$ in this case represents a symmetric object under permutations. In this way we obtain the new integrals of motion of the Hamiltonian $H_g$,
  \begin{align}\label{intcal}
   A_g&=A_F+2g(g-1)\left(\frac{x_3^2+2x_1x_2}{(x_1-x_2)^2}+\frac{x_1^2+2x_2x_3}{(x_2-x_3)^2}+\frac{x_2^2+2x_3x_1}{(x_3-x_1)^2}\right) \, , \\
 \label{cascal}
     C_{g}&= C_{3}^{\perp} +\frac{g(g-1)}{3}\frac{(x_1+x_2-2x_3)^2(x_2+x_3-2x_1)^2(x_3+x_1-2x_2)^2}{(x_1-x_2)^2(x_2-x_3)^2(x_3-x_1)^2} \, . 
 \end{align}
By construction, the two new integrals are also invariants under the permutation $[A_g, s_{\ell m}]=[C_g, s_{\ell m}]=0$ and, together with the Hamiltonian, are the $n=3$ Liouville integrals of the extended two-center Calogero system,
\begin{equation}
\text{Liouville integrals:} \quad \left\{H_{g},A_{g}, C_{g} \right\} \, .
\end{equation}
Since the resulting integrals are second order in the momenta, it is natural to expect separability of the Schr\"odinger equation. Obviously, when $g=0$ the system reduces to the most general two-center Coulomb extension, depending only in $r_{\pm}$, which was studied in section \ref{sec4}. The case $F_{\mu}=F_{\nu}=0$, $g\neq 0$ was studied in \cite{TA, TA2}, where it was also shown that the Calogero interaction prevents the complete separability due to angular coupling and the Schr\"odinger equation is partially separated in elliptic coordinates for $n>3$. In our extended case with $n=3$ and the Calogero interaction, we can separate the equation with the same ansatz as in Eq. (\ref{ans}) $\psi(\mu,\nu,\varphi)=\chi_+(\mu)\chi_-(\nu)\Xi(\varphi)$. The equations for $\chi_+(\mu)$ and $\chi_-(\nu)$ are exactly the same as before, plugging $n=3$ in (\ref{eqs1}) and (\ref{eqs2}). The angular equation (\ref{angeq}) is given in terms of the well-known P\"osch-Teller potential,
\begin{align}\notag
\nabla^2_1 \Xi(\varphi)=-q^2 \Xi(\varphi), \quad \nabla^2_1=\frac{1}{2}\partial_\varphi^2+\frac{9}{2} \cos^{-2} (3\varphi) \, .
\end{align}
This equation and its relation to angular Calogero models had also been studied before, see for instance \cite{tav, tetra}. Thus the extended system $H_g$ is endowed with an integrable structure underlying the separability of the Schr\"odinger equation, but has no further charges to induce superintegrability.

\section{Discussion}  \label{sec6}

The two-center Coulomb provides an interesting landscape to study integrable properties and different deformations. First, it is interesting to comment about the emergence of superintegrability in the two-center case and its extensions (\ref{hamilf}). For a fixed dimension, the number of independent conserved projected angular momentum components $L_{ij}^{\perp}$ is lower than in comparison with the central problem. If the number of additional symmetries, such as the Laplace-Runge-Lenz vector or $A_F$, does not increase there should be a dimension where the conserved angular momentum $L_{ij}^{\perp}$ provides enough independent integrals. In fact, it is known that both the scalar $A_F$ and the  Laplace-Runge-Lenz vector provide only one independent charge. In the extended two-center Coulomb case this critical dimension is $n=4$, making the case $n=3$ integrable but not superintegrable.

In this paper we focus on the case of completely separable models with second order Liouville integrals. A natural question is what happens beyond separability and hence beyond quadratic charges. The Calogero models are indeed a perfect scenario to study this problem. Considering our present  three dimensions results and also \cite{TA, TA2}, the case $n>3$ should have higher order charges in terms of the operators $L_{ij}^{\perp}$ obeying symmetry rules under the corresponding Weyl reflections associated with the root system. This is fully consistent with the fact that the angular reduction of the Calogero models is no longer separable for $n \geq 4$ \cite{tetra}. If we move away from separability and quadratic charges, there are many possibilities to construct further integrable models based on Calogero interactions, studying different root systems, but also for trigonometric, hyperbolic or even elliptic interactions. These cases will be considered elsewhere.

The two-center Coulomb problem has also been studied from the approach of supersymmetric quantum mechanics \cite{gmt, susy1}. The separable extended cases studied here with the Hamiltonian (\ref{hamilf}) and (\ref{hamilcal}) are also interesting to study with this method. In the latter case one could also expect the existence of intertwining operators, which are provided for the algebraic structure in the form of Dunkl operators \cite{opdam, heckman, clp, ccl}. In the standard Calogero interaction, while the conserved charges appear as Weyl invariant polynomials in terms of Dunkl operators, the intertwining ones appear as Weyl anti-invariant polynomials, see for example \cite{ccl}. It is not clear whether it would be possible to find such antilinear operators for the extended two-center Coulomb problem.

It might also be interesting also to study the deformed Dunkl algebra for the two-center model in the same spirit as it has been studied for the Calogero model with Laplace-Runge-Lenz interaction or the connection with symplectic algebras \cite{fh,th}. Another possible compatible extension could be the integrable models embedded in different geometries, where Calogero and related problems have already been studied \cite{sphere1, sphere2}.

\section*{Acknowledgements}
FC and OQ were supported by Fondecyt grant 1211356. OQ thanks City, University of London for kind hospitality, where part of this work was carried during a visit at the Department of Mathematics. OQ would also like to thank Universidad Austral de Chile, where this work originally started.

{}

\end{document}